\documentclass[a4paper, 12pt]{article} 
\usepackage[right=1.1in,left=1.1in,top=1.1in,bottom=1.1in]{geometry}
\usepackage{graphicx} 
\usepackage{braket}
\usepackage{mathrsfs}
\usepackage[T1]{fontenc} 
\usepackage{amsmath}
\usepackage{pxfonts}
\usepackage[T1]{fontenc}
\usepackage{amssymb}
\usepackage{natbib}
\usepackage{epstopdf}
\usepackage{setspace}
\usepackage{enumitem}
\usepackage{sectsty}
\usepackage{wrapfig} 
\sectionfont{\large}
\bibpunct{(}{)}{,}{a}{}{,}
\usepackage{tikz}   
\usepackage{tikz-3dplot} 

\newcommand{\qed}{\nobreak \ifvmode \relax \else
      \ifdim\lastskip<1.5em \hskip-\lastskip
      \hskip1.5em plus0em minus0.5em \fi \nobreak
      \vrule height0.75em width0.5em depth0.25em\fi}

\usepackage{bbm}
\usepackage{MnSymbol}
\usepackage[all]{xy}

\usepackage{xcolor,colortbl,array,amssymb}

   \usepackage{braket}
   \usepackage{amssymb}
\usepackage{multicol}

\makeatletter
\newlength{\bibitemsep}
\newlength{\bibparskip}
\let\oldthebibliography\thebibliography
\renewcommand\thebibliography[1]{%
  \oldthebibliography{#1}%
  \setlength{\parskip}{\bibitemsep}%
  \setlength{\itemsep}{\bibparskip}%
}

\title{ \Large{Quantum Mechanics in a Time-Asymmetric Universe:  }
\\ \Large{On the Nature of the Initial Quantum State}
} 

\author{Eddy Keming Chen\thanks{Department of Philosophy,  University of California, San Diego, 9500 Gilman Dr, La Jolla, CA 92093-0119. Website: www.eddykemingchen.net. Email: eddykemingchen@ucsd.edu  }}

\date{\emph{The British Journal for the Philosophy of Science} \textbf{74}(4), 2021 \\ First published online on October 13, 2018} 

\begin{document}
\bibliographystyle{apalike}

\maketitle 



\begin{abstract}

In a quantum universe with a strong arrow of time, we postulate a low-entropy boundary condition (the Past Hypothesis) to account for the temporal asymmetry. In this paper, I show that the Past Hypothesis also contains enough information to simplify the quantum ontology and  define a natural initial condition. 

First, I introduce \emph{Density Matrix Realism}, the thesis that the quantum state of the universe is objective and impure. 
This stands in sharp contrast to \emph{Wave Function Realism}, the thesis that the quantum state of the universe is objective and pure.

Second, I suggest that the Past Hypothesis is sufficient to determine a natural  density matrix, which is  simple and unique. This is achieved by what I call the \emph{Initial Projection Hypothesis}: the initial density matrix of the universe is the (normalized) projection onto the Past Hypothesis subspace (in the Hilbert space).  

Third, because the initial quantum state is unique and simple,  we have a strong case for the \emph{Nomological Thesis}: the initial quantum state of the universe is on a par with laws of nature. 

This new package of ideas has several interesting implications, including on the harmony between statistical mechanics and quantum mechanics,  theoretical unity of the universe and the subsystems, and the alleged conflict between Humean supervenience and quantum entanglement. 
 
\end{abstract}

\hspace*{3,6mm}\textit{Keywords: quantum state of the universe, time's arrow, Past Hypothesis, Statistical Postulate, the Mentaculus Vision, Wentaculus, typicality, unification, foundations of probability, quantum statistical mechanics, wave function realism, quantum ontology, density matrix, Weyl Curvature Hypothesis, Humean Supervenience}   

\newpage

\begingroup
\singlespacing
\tableofcontents
\endgroup

\vspace{20pt} 



\nocite{ AlbertLPT, loewer2004david, lebowitz2008time, goldstein2001boltzmann, durr1992quantum, durr2012quantum, goldstein2013reality, goldstein2010approachB, goldstein2010approach, goldstein2010normal, durr2005role, bell1980broglie, goldstein2012typicality, ney2013wave, ChenOurFund, ChenHSWF, chen2017intrinsic, allori2013primitive, allori2008common, loewer2016mentaculus, LewisPP2, sep-time-thermo, north2011time, coen2010serious, emery2017against, lewis2016quantum, monton2006quantum}

\section{Introduction}

In the foundations of quantum mechanics, it has been argued that the wave function (pure state) of the universe represents something objective and not something merely epistemic. Let us call this view \emph{Wave Function Realism}. There are many realist proposals for how to understand the wave function. Some argue that it represents things in the ontology, either a physical field propagating on a fundamental high-dimensional space, or a multi-field propagating on the three-dimensional physical space. Others argue that it is in the ``nomology''---having the same status as laws of nature. Still others argue that it might belong to a new ontological category.\footnote{See \cite{AlbertEQM, LoewerHS, wallace2010quantum, NeySOTDQU, NorthSQW, maudlin2013nature, goldstein2013reality, miller2014quantum, esfeld2014quantum, bhogal2015humean, callender2015one, esfeld2017minimalist, chen2017intrinsic, ChenOurFund, chen2016, Hubert2018}. For a survey of this literature, see \cite{chen2019realism}. Notice that this is not how Albert, Loewer, or Ney characterizes wave function realism. For them, to be a wave function realist is to be a realist about the wave function and a fundamental high-dimensional space---the ``configuration space.'' For the purpose of this paper, let us use \emph{Wave Function Realism} to designate just the commitment that the wave function represents something objective.} 

Wave Function Realism has generated much debate. In fact, it has been rejected by many people, notably by quantum Bayesians, and various anti-realists and instrumentalists. As a scientific realist, I do not find their arguments convincing. In previous papers, I have assumed and defended Wave Function Realism. Nevertheless, in this paper I want to argue for a different perspective, for  reasons  related to the origin of time-asymmetry in a quantum universe. 

To be sure, realism about the universal wave function is highly natural in the context of standard quantum mechanics and various realist quantum theories such as Bohmian mechanics (BM), GRW spontaneous collapse theories, and Everettian quantum mechanics (EQM). In those theories, the universal wave function is indispensable to the kinematics and the dynamics of the quantum system. However,  as I would like to emphasize in this paper, our world is not just quantum-mechanical. We also live in a world with a strong arrow of time (large entropy gradient). There are  thermodynamic phenomena that we hope to explain with quantum mechanics and quantum statistical mechanics. A central theme of this paper is to suggest that quantum statistical mechanics is highly relevant for assessing the fundamentality and  reality of the universal wave function. 

We will take a close look at the connections between the foundations of quantum statistical mechanics and various solutions to the quantum measurement problem. When we do, we realize that we do not \emph{need} to postulate a universal wave function. We  need only certain ``coarse-grained'' information about the quantum macrostate, which can be represented by a density matrix. A natural question is: can we understand the universal quantum state as a density matrix rather than a wave function? That is, can we take an ``ontic'' rather than an ``epistemic'' attitude towards the density matrix? 

The first step of  this paper is to argue that we can. I call this view \emph{Density Matrix Realism}, the thesis that the actual quantum state of the universe is objective (as opposed to subjective or epistemic) and impure (mixed). 
This idea may be unfamiliar to some people, as we are used to take the mixed states to represent our \emph{epistemic uncertainties of the actual pure state} (a wave function).  The proposal here is that the density matrix directly represents the actual quantum state of the universe; there is no further fact about which is the actual wave function. In this sense, the density matrix is ``fundamental.'' In fact,  this idea has come up  in the foundations of physics.\footnote{See, for example, \cite{durr2005role, maroney2005density}, \cite{wallace2011logic, wallace2012emergent},  and \cite{wallace2016probability}.} In the first step, we provide a systematic discussion of Density Matrix Realism by reformulating Bohmian mechanics, GRW theories, and Everettian quantum mechanics in terms of a \emph{fundamental} density matrix. 

The second step is to point out that Density Matrix Realism allows us to combine quantum ontology with time-asymmetry in a new way. In classical and quantum statistical mechanics,  thermodynamic time-asymmetry arises from a special boundary condition called the Past Hypothesis.\footnote{For an extended discussion, see \cite{albert2000time}.} I suggest that the information in the \emph{Past Hypothesis}  is sufficient to determine a natural  density matrix. I postulate the \emph{Initial Projection Hypothesis}: the quantum state of the universe at $t_0$ is given by the (normalized) projection onto the Past Hypothesis subspace, which is a particular low-dimensional subspace in the total Hilbert space. 
The conjunction of this hypothesis with Density Matrix Realism pins down a \emph{unique} initial quantum state. Since the Initial Projection Hypothesis is as simple as the Past Hypothesis, we can use arguments for the simplicity of the latter (which is necessary for it to be a law of nature) to argue for the simplicity of the former. We can thus infer that the initial quantum state is very \emph{simple}. 

The third step is to show that, because of the simplicity and the uniqueness of the initial quantum state (now given by a fundamental density matrix), we have a strong case for the \emph{Nomological Thesis}: the initial quantum state of the world is on a par with laws of nature. It is a modal thesis. It implies that the initial quantum state of our world is nomologically necessary; it could not have been otherwise. 

As we shall see, this package of views has interesting implications for the reduction of statistical mechanical probabilities to quantum mechanics, the dynamic and kinematic unity of the universe and the subsystems, the nature of the initial quantum state, and Humean supervenience in a quantum world. 

Here is the roadmap of the paper. First, in \S 2, I review the foundations of quantum mechanics and quantum statistical mechanics. In \S 3, I introduce the framework of Density Matrix Realism and provide some illustrations. In \S 4, I propose the Initial Projection Hypothesis in the framework of  Density Matrix Realism. In \S 5, I discuss their implications for statistical mechanics, dynamic unity, and kinematic unity. In \S 6, I suggest that they provide a strong case for the Nomological Thesis and a new solution to the conflict between quantum entanglement and Humean supervenience. 

\section{Foundations of Quantum Mechanics and Statistical Mechanics}

In this section, we first review the foundations of quantum mechanics and statistical mechanics. As we shall see in the next section, they suggest an alternative to Wave Function Realism. 

\subsection{Quantum Mechanics}

Standard quantum mechanics is often presented with a set of axioms and rules about measurement. Firstly, there is a quantum state of the system, represented by a wave function $\psi$. For a spin-less $N$-particle quantum system in $\mathbb{R}^3$, the wave function is a (square-integrable) function from the configuration space $\mathbb{R}^{3N}$ to the complex numbers $\mathbb{C}$. Secondly, the wave function evolves in time according to the the Schr\"odinger equation:

\begin{equation}\label{SE}
 i\hbar \frac{\partial \psi}{\partial t} = H \psi
\end{equation}
Thirdly, the Schr\"odinger evolution of the wave function is supplemented with collapse rules. The wave function typically evolves into superpositions of macrostates, such as the cat being alive and the cat being dead. This can be represented by wave functions on the configuration space with disjoint macroscopic supports $X$ and $Y$. During measurements, which are not precisely defined processes in the standard formalism, the wave function  undergoes collapses. Moreover,  the probability that it collapses into any particular macrostate $X$ is given by the Born rule: 

\begin{equation}\label{Born}
P( X) = \int_{X} |\psi(x)|^2 dx
\end{equation}

As such, quantum mechanics is not a candidate for a fundamental physical theory. It has two dynamical laws: the deterministic Schr\"odinger equation and the stochastic collapse rule. What are the conditions for applying the former, and what are the conditions for applying the latter? Measurements and observations are extremely vague concepts. Take a concrete experimental apparatus for example. When should we treat it as part of the quantum system that evolves linearly and when should we treat it as an ``observer,'' i.e. something that stands outside the quantum system and collapses the wave function? That is, in short, the quantum measurement problem.\footnote{See \cite{bell1990against} and \cite{sep-qt-issues} for  introductions to the quantum measurement problem.} 

Various solutions have been proposed regarding the measurement problem. Bohmian mechanics (BM) solves it by adding particles to the ontology and an additional guidance equation for the particles' motion. Ghirardi-Rimini-Weber (GRW)  theories postulate a spontaneous collapse mechanism. Everettian quantum mechanics (EQM) simply removes the collapse rules from standard quantum mechanics and suggest that there are many (emergent) worlds, corresponding to the branches of the wave function, which are all real. My aim here is not to adjudicate among these theories. Suffice it to say that they are all quantum theories that remove the centrality of observations and observers. 


To simplify the discussions, I will use BM as a key example.\footnote{See \cite{durr1992quantum} for a rigorous presentation of BM and its statistical analysis.} In BM, in addition to the wave function that evolves unitarily according to the Schr\"odinger equation,   particles have precise locations, and their configuration $Q = (Q_1, Q_2, ... , Q_N)$ follows the guidance equation: 
\begin{equation}\label{GE}
 \frac{dQ_i}{dt} = \frac{\hbar}{m_i} \text{Im} \frac{ \nabla_i \psi (q) }{  \psi (q)} (q=Q)
\end{equation}
Moreover, the initial particle distribution is given by the quantum equilibrium distribution: 
\begin{equation}\label{QEH}
\rho_{t_0} (q) = |\psi(q, t_0)|^2
\end{equation}
By equivariance, if this condition holds at the initial time, then it holds at all times. Consequently, BM agrees with standard quantum mechanics with respect to the Born rule predictions (which are all there is to the observable predictions of quantum mechanics).  For a universe with $N$ particles, let us call the wave function of the universe the \emph{universal wave function} and denote it by $\Psi(\boldsymbol{q_1, q_2, ... q_N})$. 

\subsection{Quantum Statistical Mechanics}

Statistical mechanics concerns macroscopic systems such as gas in a box. It is an important subject for  understanding the arrow of time. For concreteness, let us consider a quantum-mechanical system with $N$ fermions (with $N> 10^{20}$) in a box $\Lambda = [0, L]^3 \subset \mathbb{R}^3$ and a Hamiltonian $\hat{H}$. I will first present the ``individualistic'' view followed by the ``ensemblist'' view of quantum statistical mechanics (QSM).\footnote{Here I follow the discussions in \cite{goldstein2010approach} and \cite{goldstein2010approachB}.} I will include some brief remarks comparing  QSM to classical statistical mechanics (CSM), the latter of which may be more familiar to some readers.

\begin{enumerate}
\item Microstate: at any time $t$, the microstate of the system is given by a normalized (and anti-symmetrized) wave function:
\begin{equation}
\psi(\boldsymbol{q_1}, ..., \boldsymbol{q_N}) \in \mathscr{H}_{total} = L^2 (\mathbb{R}^{3N}, \mathbb{C}^k) \text{ , } \parallel \psi \parallel_{L^2} = 1,
\end{equation}
where $\mathscr{H}_{total} = L^2 (\mathbb{R}^{3N}, \mathbb{C}^k)$ is the total Hilbert space of the system. (In CSM, the microstate is given by the positions and the momenta of all the particles, represented by a point in phase space.) 

\item Dynamics: the time dependence of $\psi(\boldsymbol{q_1}, ..., \boldsymbol{q_N}; t)$ is given by the Schr\"odinger equation: 
\begin{equation}
 i\hbar \frac{\partial \psi}{\partial t} = H \psi.
\end{equation}
(In CSM, the particles move according to the Hamiltonian equations.)

\item Energy shell: the physically relevant part of the total Hilbert space is the subspace (``the energy shell''):
\begin{equation}
\mathscr{H} \subseteq \mathscr{H}_{total} \text{ , } \mathscr{H} = \text{span} \{ \phi_\alpha : E_\alpha \in [E, E+\delta E ]  \},
\end{equation}
This is the subspace (of the total Hilbert space) spanned by energy eigenstates $\phi_\alpha$ whose eigenvalues $E_\alpha$ belong to the $[E, E+\delta E]$ range.  Let $D = \text{dim} \mathscr{H}$, the number of energy levels between $E$ and $E+\delta E$. 

We only consider wave functions $\psi$ in $\mathscr{H}$.

\item Measure: the measure $\mu$ is given by the normalized surface area measure on the unit sphere in the energy subspace $\mathscr{S}(\mathscr{H})$.

\item Macrostate: with a choice of macro-variables (suitably ``rounded'' \emph{\`a la} \cite{von1955mathematical}), the energy shell $\mathscr{H}$ can be orthogonally decomposed into macro-spaces:
\begin{equation}
\mathscr{H} = \oplus_\nu \mathscr{H}_\nu \text{ , } \sum_\nu \text{dim}\mathscr{H}_\nu  = D
\end{equation}
Each $\mathscr{H}_\nu$ corresponds more or less to small ranges of values of macro-variables that we have chosen in advance. (In CSM, the phase space can be partitioned into sets of phase points. They will be the macrostates.) 

\item Non-unique correspondence: typically, a wave function is in a superposition of macrostates and is not entirely in any one of the macrospaces. However, we can make sense of situations where $\psi$ is (in the Hilbert space norm) very close to a macrostate $\mathscr{H}_\nu$: 
\begin{equation}\label{close}
\bra{\psi} P_{\nu}  \ket{\psi} \approx 1,
\end{equation}
where $P_{\nu}$ is the projection operator onto $\mathscr{H}_{\nu}$. This means that almost all of $\ket{\psi}$ lies in $\mathscr{H}_{\nu}$. (In CSM, a phase point is always entirely within some macrostate.)

\item Thermal equilibrium: typically, there is a dominant macro-space $\mathscr{H}_{eq}$ that has a dimension that  is almost equal to D: 
\begin{equation}
\frac{\text{dim} \mathscr{H}_{eq}}{\text{dim} \mathscr{H}} \approx 1.
\end{equation}
A system with wave function $\psi$ is in equilibrium if the wave function $\psi $ is very close to $\mathscr{H}_{eq}$ in the sense of (\ref{close}):  $\bra{\psi} P_{eq}  \ket{\psi} \approx 1.$

\emph{Simple Example.} Consider a gas consisting of $n = 10^{23}$ atoms in a box $\Lambda \subseteq \mathbb{R}^3.$ The system is governed by quantum mechanics. We orthogonally decompose the Hilbert space $\mathscr{H}$ into 51 macro-spaces: $\mathscr{H}_0 \oplus \mathscr{H}_2  \oplus \mathscr{H}_4  \oplus...  \oplus \mathscr{H}_{100},$ where $\mathscr{H}_\nu$ is the subspace corresponding to the macrostate such that the number of atoms in the left half of the box is between $(\nu-1) \%$ and $(\nu+1) \%$ of $n$, with the endpoints being the exceptions: $\mathscr{H}_{0}$ is the interval $0\%-1\%$, and $\mathscr{H}_{100}$ is the interval $99\%-100\%$. In this example, $\mathscr{H}_{50}$ has the overwhelming majority of dimensions and is thus the equilibrium macro-space. A system whose wave function is very close to $\mathscr{H}_{50}$ is in equilibrium (for this choice of macrostates).

\item Boltzmann Entropy: the Boltzmann entropy of a quantum-mechanical system with wave function $\psi$ that is very close to a macrostate $\mathscr{H}_\nu$ is given by:
\begin{equation}\label{Boltzmann}
S_B (\psi) = k_B \text{log} (\text{dim} \mathscr{H}_\nu ),
\end{equation}
where $\mathscr{H}_\nu$ denotes the subspace containing almost all of $\psi$ in the sense of (\ref{close}). The thermal equilibrium state thus has the maximum entropy: 
\begin{equation}
S_B (eq) = k_B \text{log} (\text{dim} \mathscr{H}_{eq} ) \approx  k_B \text{log} (D),
\end{equation}
where $\mathscr{H}_{eq}$ denotes the equilibrium macrostate. (In CSM, Boltzmann entropy of a phase point is proportional to the logarithm of the volume measure of the macrostate it belongs to.)

\item Low-Entropy Initial Condition: when we consider the universe as a quantum-mechanical system, we postulate a special low-entropy boundary condition on the universal wave function---the quantum-mechanical version of the \emph{Past Hypothesis}: 
\begin{equation}
\Psi(t_0) \in \mathscr{H}_{PH} \text{ , } \text{dim} \mathscr{H}_{PH} \ll \text{dim}\mathscr{H}_{eq} \approx \text{dim} \mathscr{H}
\end{equation}
where $\mathscr{H}_{PH}$ is the Past Hypothesis macro-space with dimension much smaller than that of the equilibrium macro-space.\footnote{We should assume that $\mathscr{H}_{PH}$ is finite-dimensional, in which case we can use the normalized surface area measure on the unit sphere as the typicality measure for \# 10. It remains an open question in QSM about how to formulate the low-entropy initial condition when the initial macro-space is infinite-dimensional.} Hence, the initial state has very low entropy in the sense of (\ref{Boltzmann}). (In CSM, the Past Hypothesis says that the initial microstate is in a low-entropy macrostate with very small volume.)

\item A central task of QSM is to establish mathematical results that demonstrate (or suggest) that for $\mu-$most (maybe even all) wave functions, the small subsystems, such as gas in a box,  will approach thermal equilibrium. 
\end{enumerate}

Above is the individualistic view of QSM in a nutshell. In contrast, the ensemblist view of QSM differs in several ways. First, on the ensemblist view, instead of focusing on the wave function of an individual system, the focus is on an ensemble of systems that have the same statistical state $\hat{W}$, a density matrix.\footnote{Ensemblists would further insist that it makes no sense to talk about the thermodynamic state of an individual system.} 
It evolves according to the von Neumann equation:
\begin{equation}\label{VNM}
i \hbar \frac{d \hat{W}(t)}{d t} = [\hat{H},  \hat{W}].
\end{equation}

The crucial difference between the individualistic and the ensemblist views of QSM lies in the definition of thermal equilibrium. On the ensemblist view, a system is in thermal equilibrium if:
\begin{equation}
W = \rho_{mc} \text{ or } W = \rho_{can},
\end{equation}
where $\rho_{mc}$ is the microcanonical ensemble and $\rho_{can}$ is the canonical ensemble.\footnote{The microcanonical ensemble is the projection operator onto the energy shell  $\mathscr{H}$ normalized by its dimension. The canonical ensemble is:  
\begin{equation}\label{canonical}
\rho_{can} = \frac{\text{exp}(-\beta \hat{H})}{Z},
\end{equation}
where $Z = \text{tr } \text{exp}(-\beta \hat{H})$, and $\beta$ is the inverse temperature of the quantum system.
}

For the QSM individualist, if the microstate $\psi$ of a system is close to some macro-space $\mathscr{H}_{\nu}$ in the sense of (\ref{close}), we can say that the macrostate of the system is $\mathscr{H}_{\nu}$. The macrostate is naturally associated with a density matrix: 
\begin{equation}\label{ID}
\hat{W}_{\nu} = \frac{I_{\nu}}{dim \mathscr{H}_{\nu}},
\end{equation}
where $I_{\nu}$ is the projection operator onto  $\mathscr{H}_{\nu}$. $\hat{W}_{\nu}$ is thus a representation of the macrostate. It can be decomposed into wave functions, but the decomposition is not unique. Different measures can give rise to the same density matrix.  One such choice is $\mu(d\psi)$, the uniform distribution over wave functions: 
\begin{equation}\label{MacroW}
\hat{W}_{\nu} = \int_{\mathscr{S}(\mathscr{H_{\nu}})} \mu(d\psi) \ket{\psi} \bra{\psi}.
\end{equation}
In (\ref{MacroW}), $\hat{W}_{\nu}$ is defined with a choice of measure on wave functions in $\mathscr{H_{\nu}}$. However, we should not be misled into thinking that the density matrix is derived from wave functions. What is  intrinsic to a density matrix is its geometrical meaning in the Hilbert space. In the case of $\hat{W}_{\nu}$,  as shown in the canonical description (\ref{ID}), it is just a normalized projection operator.\footnote{Thanks to Sheldon Goldstein for helping me appreciate the intrinsic meaning of density matrices. That was instrumental in the final formulation of the Initial Projection Hypothesis in \S 4.2.} 

\section{Density Matrix Realism}

According to Wave Function Realism,  the quantum state of the universe is objective and pure. On this view, $\Psi$ is both the microstate of QSM and a dynamical object of QM. 

Let us recall the arguments for Wave Function Realism. Why do we attribute objective status to the quantum state represented by a wave function?  It is because the wave function plays crucial roles in the realist quantum theories. In BM, the wave function appears in the fundamental dynamical equations and guides particle motion. In GRW, the wave function spontaneously collapses and gives rise to macroscopic configurations of tables and chairs. In EQM, the wave function is the whole world. If the universe is accurately described by BM, GRW, or EQM, then the wave function is an active ``agent'' that makes a difference in the world.  The wave function cannot  represent just our ignorance. It has to be objective, so the arguments go. But what is the nature of the quantum state that it represents? As mentioned in the beginning of this paper, there are several interpretations: the two field interpretations, the nomological interpretation, and the \emph{sui generis} interpretation. 

On the other hand, we often use $W$, a density matrix, to represent our ignorance of $\psi$, the actual wave function of a quantum system.  $W$  can also represent a macrostate in QSM.\footnote{In some cases, $W$ is  easier for calculation than $\Psi$, such as in the case of GRW collapse theories where there are multiple sources of randomness. Thanks to Roderich Tumulka for discussions here. }

Is it possible to be a realist about the density matrix of the universe and attribute objective status to the quantum state it represents? That depends on whether we can write down realist quantum theories directly in terms of $W$. Perhaps $W$ does not have enough information to be the basis of a realist quantum theory. However, if we can formulate quantum dynamics directly in terms of $W$ instead of $\Psi$ such that $W$ guides Bohmian particles, or $W$ collapses, or $W$ realizes the emergent multiverse, then we will have good reasons for taking $W$ to represent something objective in those theories. At the very least, the reasons for that will be on a par with those for Wave Function Realism in the $\Psi$-theories. 

However, can we describe the quantum universe with $W$ instead of $\Psi$? The answer is yes.  \cite{durr2005role} has worked out the Bohmian version. In this section, I describe how. Let us call this new framework \emph{Density Matrix Realism}.\footnote{The possibility that the universe can be described by a fundamental density matrix (mixed state) has been suggested by multiple authors and explored to various extents (see Footnote \#2).  What is new in this paper is the combination of Density Matrix Realism with the Initial Projection Hypothesis (\S 4) and the argument for the Nomological Thesis (\S 6) based on that. However, Density Matrix Realism is unfamiliar enough to warrant some clarifications and developments.}  I will use W-Bohmian Mechanics as the main example and explain how a fundamental density matrix can be empirically adequate for describing a quantum world. We can also construct  W-Everett theories and W-GRW theories. Similar to Wave Function Realism, Density Matrix Realism is open to several interpretations. At the end of this section, I will  provide three field interpretations of $W$. In \S 6, I discuss and motivate a nomological interpretation.

\subsection{W-Bohmian Mechanics}

First, we  illustrate the differences between Wave Function Realism and Density Matrix Realism by thinking about two different Bohmian theories. 

In standard Bohmian mechanics (BM), an $N$-particle universe at a time $t$  is  described by ($Q(t)$, $\Psi(t)$). The universal wave function guides particle motion and provides the probability distribution of particle configurations. Given the centrality of $\Psi$ in BM, we take the wave function to represent something objective (and it is open to several realist interpretations). 

It is somewhat surprising that we can formulate a Bohmian theory with only $W$ and $Q$. This was  introduced as W-Bohmian Mechanics (W-BM) in \cite{durr2005role}. The fundamental density matrix $W(t)$ is governed by the von Neumann equation (\ref{VNM}). Next, the particle configuration $Q(t)$ evolves according to an analogue of the guidance equation (W-guidance equation):  

\begin{equation}\label{WGE}
\frac{dQ_i}{dt} = \frac{\hbar}{m_i} \text{Im} \frac{\nabla_{q_{i}}  W (q, q', t)}{ W (q, q', t)} (q=q'=Q),
\end{equation}
(Here we have set aside spin degrees of freedom. If we include spin, we can add the partial trace operator $\text{tr}_{\mathbb{C}^k}$ before each occurrence of ``$W$.'')
Finally, we can  impose an initial probability distribution  similar to that of the quantum equilibrium distribution: 
\begin{equation} \label{WQEH}
P(Q(t_0) \in dq) = W (q, q, t_0) dq.
\end{equation}
The system is also equivariant: if the probability distribution holds at $t_0$, it holds at all times.\footnote{Equivariance holds because of the following continuity equation: 
$$\frac{\partial  W(q,q,t) }{\partial t} = -\text{div} (  W(q, q, t) v),$$ where $v$ denotes the velocity field generated via (\ref{WGE}). See \cite{durr1992quantum, durr2005role}.} 

With the defining equations---the von Neumann equation (\ref{VNM}) and the W-guidance equation (\ref{WGE})---and the initial probability distribution (\ref{WQEH}), we have a theory that directly uses a  density matrix $W(t)$ to characterize the trajectories $Q(t)$ of the universe's $N$ particles. If a universe is accurately described by W-BM, then $W$ represents the fundamental quantum state in the theory that guides particle motion; it does not do so via some other entity $\Psi$. If we have good reasons to be a wave function realist in BM, then we have equally good reasons to be a density matrix realist in W-BM. 

W-BM is empirically equivalent to BM with respect to the observable quantum phenomena, that is, pointer readings in quantum-mechanical experiments. By the usual typicality analysis (\cite{durr1992quantum}), this follows from (\ref{WQEH}), which is analogous to  the quantum equilibrium distribution in BM. With the respective dynamical equations, both BM and W-BM generate an equivariant Born-rule probability distribution over all measurement outcomes.\footnote{Here I am assuming that two theories are empirically equivalent  if they assign the same probability distribution to all possible outcomes of experiments. This is the criterion used in the standard Bohmian statistical analysis (\cite{durr1992quantum}). Empirical equivalence between BM and W-BM follows from the equivariance property plus the quantum equilibrium distribution. Suppose W-BM is governed by a universal density matrix $W$ and suppose BM is governed by a universal wave function chosen at random whose statistical density matrix is $W$. Then the initial particle distributions on both theories are the same: $W(q, q, t_0)$. By equivariance, the particle distributions will always be the same. Hence, they always agree on what is typical. See \cite{durr2005role}.
This is a general argument. In \cite{chen2019quantum1}, I present the general argument followed by a subsystem  analysis of W-BM, in terms of conditional density matrices.}

\subsection{W-Everettian and W-GRW Theories}

W-BM is a  quantum theory in which the density matrix is objective. In this theory, realism about the universal density matrix is based on the central role it plays in the laws of a W-Bohmian universe: it appears in the fundamental dynamical equations and it guides particle motion. (In \S 3.3, we will provide three concrete physical interpretations of $W$.) What about other quantum theories, such as Everettian and GRW theories? Is it possible to ``replace'' their universal wave functions with  universal density matrices? We  show that such suggestions are also possible.\footnote{Thanks to Roderich Tumulka, Sheldon Goldstein, and Matthias Lienert for discussions here. The W-GRW formalism was suggested first in \cite{allori2013predictions}.}    First, let us define \emph{local beables} (\`a la \cite{bell2004speakable}). Local beables are the part of the ontology that is localized (to some bounded region) in physical space. Neither the total energy function nor the wave function is a local beable. Candidate local beables include particles, space-time events (flashes), and matter density ($m(x,t)$).

For the Everettian theory with no local beables (S0), we can postulate that the fundamental quantum state is represented by a density matrix $W(t)$ that evolves unitarily by the von Neumann equation (\ref{VNM}). Let us call this theory W-Everett theory (W-S0). Since there are no additional variables in the theory, the density matrix represents the entire quantum universe. The density matrix will give rise to many branches that (for all practical purposes) do not interfere with each other.  The difference is that there will be (in some sense) more branches in the W-Everett quantum state than in the Everett quantum state. In the W-Everett universe, the world history will be described by the undulation of the density matrix.\footnote{W-S0 is a novel version of Everettian theory, one that will require more mathematical analysis to fully justify the emergence of macroscopic branching structure. It faces the familiar preferred-basis problem as standard Everett does. In addition, on W-S0 there will be some non-uniqueness in the decompositions of the Hilbert space into macrospaces. I leave the analysis for future work.}

It is difficult to find tables and chairs in a universe described only by a quantum state. One proposal is to add ``local beables'' to the theory in the form of a mass-density ontology $m(x,t)$. The wave-function version was introduced as Sm by \cite{allori2010many}. The idea is that the wave function evolves by the Schr\"odinger equation and determines the shape of the mass density. This idea can be used to construct a density-matrix version (W-Sm).  In this theory, $W(t)$ will evolve unitarily by the von Neumann equation. Next, we can define the mass-density function directly in terms of $W(t)$:
\begin{equation}\label{mxt}
m(x,t) = \text{tr} (M(x) W(t)),
\end{equation}
where $x$ is a physical space variable, $M(x) = \sum_i m_i \delta (Q_i - x)$ is the mass-density operator, which is defined via the position operator $Q_i \psi (q_1, q_2, ... q_n)= q_i \psi (q_1, q_2, ... q_n) $. This allows us to determine the mass-density ontology at time $t$  via $W(t)$.

For the density-matrix version of GRW theory with just a quantum state (W-GRW0), we  need to introduce the collapse of a density matrix. Similar to the wave function in GRW0, between collapses, the density matrix in W-GRW0 will evolve unitarily according to the von Neumann equation. It collapses randomly, where the random time for an $N$-particle system is distributed with rate $N\lambda$, where $\lambda$ is of order $10^{-15}$ s$^{-1}$. At a random time when a collapse occur at ``particle'' $k$ at time $T^-$, the post-collapse density matrix at time $T^+$ is the following:
\begin{equation}\label{collapse}
W_{T^+} = \frac{\Lambda_k (X)^{1/2} W_{T^-} \Lambda_k (X)^{1/2}}{\text{tr} (W_{T^-} \Lambda_k (X)) },
\end{equation}
with $X$ distributed by the following probability density:
\begin{equation}\label{center}
\rho(x) = \text{tr} (W_{T^-} \Lambda_k (x)), 
\end{equation}
where $W_{T^+}$ is the post-collapse density matrix, $W_{T^-}$ is the pre-collapse density matrix, $X$ is the center of the actual collapse, and $\Lambda_k (x)$ is the collapse rate operator.\footnote{A collapse rate operator is defined as follows:
$$\Lambda_k (x) = \frac{1}{(2\pi \sigma^2)^{3/2}} e^{-\frac{(Q_k -x)^2}{2\sigma^2}},$$ where $Q_k$ is the position operator of ``particle'' $k$, and $\sigma$ is a new constant of nature of order $10^{-7}$ m postulated in current GRW theories. Compare W-GRW to $\Psi$-GRW, where  collapses happen at the same rate, and the post-collapse wave function is the following:
\begin{equation}\label{WFcollapse}
\Psi_{T^+} = \frac{\Lambda_k (X)^{1/2} \Psi_{T^-} }{||  \Lambda_k (X)^{1/2} \Psi_{T^-}  || },
\end{equation}
with the collapse center $X$ being chosen randomly with probability distribution $\rho(x) = ||  \Lambda_k (x)^{1/2} \Psi_{T^-}  ||^2 dx$. 
}

For the GRW theory (W-GRWm) with both a quantum state $W(t)$ and a mass-density ontology $m(x,t)$, we can combine the above  steps:  $W(t)$ evolves by the von Neumann equation that is randomly interrupted by collapses (\ref{collapse}) and $m(x,t)$ is defined by (\ref{mxt}). We can define GRW with a flash-ontology (W-GRWf) in a similar way, by using $W(t)$ to characterize the distribution of flashes  in physical space-time. The flashes are the space-time events at the centers ($X$) of the W-GRW collapses. 

To sum up: in W-S0, the entire world history is described by $W(t)$; in W-Sm, the local beables (mass-density) is determined by $W(t)$; in W-GRW theories, $W(t)$ spontaneously collapses. These roles were originally played by $\Psi$, and now they are played by W. In so far as we have good reasons for Wave Function Realism based on the roles that $\Psi$ plays in the $\Psi$-theories, we have equally good reasons for Density Matrix Realism if the universe is accurately described by  W-theories.

\subsection{Field Intepretations of $W$}

Realism about the density matrix only implies that it is objective and not epistemic. Realism is compatible with a wide range of concrete interpretations of what the density matrix represents. In this section, I provide three field interpretations of the density matrix. But they do not exhaust all available options. In \S 6, I motivate a nomological interpretation of the density matrix that is also realist. 

In debates about the metaphysics of the wave function, realists have offered several interpretations of $\Psi$. Wave function realists, such as Albert and Loewer, have offered a concrete physical interpretation: $\Psi$ represents a physical field on the high-dimensional configuration space that is taken to be the fundamental physical space.\footnote{In \cite{ChenOurFund}, I argue against this view and suggest that there are many good reasons---internal and external to quantum mechanics---for taking the low-dimensional physical space-time to be fundamental. }  

Can we interpret the density matrix in a similar way? Let us start with a mathematical representation of the density matrix $W(t)$. It is defined as a positive, bounded, self-adjoint operator $\hat{W}: \mathscr{H} \rightarrow \mathscr{H}$ with $\text{tr} \hat{W} = 1$. For W-BM, the configuration space $\mathbb{R}^{3N}$, and a density operator $\hat{W}$, the relevant Hilbert space is $\mathscr{H}$, which is a subspace of the total Hilbert space, i.e. $\mathscr{H} \subseteq \mathscr{H}_{total} =  L^2 (\mathbb{R}^{3N}, \mathbb{C})$. Now, the density matrix $\hat{W}$ can also be represented as a function 
\begin{equation}
W: \mathbb{R}^{3N} \times \mathbb{R}^{3N} \rightarrow \mathbb{C}
\end{equation}
(If we include spin, the range will be the endomorphism space $\text{End}(\mathbb{C}^k) $ of the space of linear maps from $\mathbb{C}^k$ to itself. Notice that we have already used the position representation in (\ref{WGE}) and (\ref{WQEH}).)  

This representation enables three field  interpretations of the density matrix. Let us use W-BM as an example. First, the fundamental space is represented by $\mathbb{R}^{6N}$, and $W$ represents a field on that space that assigns properties (represented by complex numbers) to each point in $\mathbb{R}^{6N}$. In the Bohmian version, $W$ guides the motion of a ``world particle'' like a river guides the motion of a ping pong ball. (However, the world particle only moves in a $\mathbb{R}^{3N}$ subspace.) Second, the fundamental space is $\mathbb{R}^{3N}$, and $W$ represents a multi-field on that space that assigns properties to every ordered pair of points $(q, q')$ in $\mathbb{R}^{3N}$. The world particle moves according to the gradient taken with respect to the first variable of the multi-field. Third, the fundamental space is the physical space represented by $\mathbb{R}^{3}$, and the density matrix represents a multi-field that assigns properties to every ordered pair of $N$-regions, where each $N$-region is composed of $N$ points in physical space. On this view, the density matrix guides the motion of $N$ particles in physical space.\footnote{For discussions about the multi-field interpretation, see \cite{forrest1988quantum, belot2012quantum}, Chen (2017), Chen (ms.) section 3, and \cite{Hubert2018}.} 

These three field interpretations are available to the  density matrix realists. In so far as we have good grounds for accepting the field interpretations of wave function realism, we have equally good grounds for accepting these interpretations for the W-theories. These physical interpretations, I hope, can provide further  reasons for  wave function realists to take seriously the idea that density matrices \emph{can} represent something physically significant. In \S 6, we  introduce a new interpretation of $W$ as something nomological, and we will motivate that with the new Initial Projection Hypothesis.  That, I believe, is the most interesting realist interpretation of the universal density matrix all things considered.

\section{The Initial Projection Hypothesis}


W-quantum theories are alternatives to $\Psi$-quantum theories. However, all of these theories are time-symmetric, as they obey time-reversal invariance. 

In statistical mechanics, a fundamental postulate is added to the time-symmetric dynamics: the Past Hypothesis, which is a low-entropy boundary condition of the universe. In this section, we will first discuss the wave-function version of the Past Hypothesis. Then we will use it to pick out a special density matrix. I call this the \emph{Initial Projection Hypothesis}. Finally, we point out some connections between the Initial Projection Hypothesis  and Penrose's Weyl Curvature Hypothesis.

\subsection{The Past Hypothesis}

The history of the Past Hypothesis goes back to Ludwig Boltzmann.\footnote{For an extended discussion, see \cite{boltzmann2012lectures}, \cite{albert2000time}, and \cite{sep-time-thermo}.} To explain time asymmetry in a universe governed by time-symmetric equations, Botlzmann's solution is to add a boundary condition: the universe started in a special state of very low-entropy.  Richard Feynman agrees, ``For some reason, the universe at one time had a very low entropy for its energy content, and since then the entropy has increased.''\footnote{\cite{feynman2015feynman}, 46-8. } Such a low-entropy initial condition  explains the arrow of time in thermodynamics.\footnote{See \cite{lebowitz2008time, ehrenfest2002conceptual} and \cite{penrose1979singularities} for more discussions about a low-entropy initial condition. See \cite{earman2006past} for worries about  the Past Hypothesis as an initial condition for the universe. See \cite{goldstein2016hypothesis} for a discussion about the possibility, and some recent examples, of explaining the arrow of time without the Past Hypothesis.}  

 Albert (2000) has called this condition the \emph{Past Hypothesis} (PH). However, his proposal is stronger than the usual one concerning a low-entropy initial condition. The usual one just postulates that the universe started in some low-entropy macrostate. It can be any of the many macrostates, so long as it has sufficiently low entropy. Albert's PH postulates that there is a \emph{particular} low-entropy macrostate that the universe starts in---the one that underlies the reliability of our inferences to the past. It is the task of cosmology to discover that initial macrostate. In what follows, I refer to the strong version of PH unless indicated otherwise.\footnote{In  \cite{chen2018valia}, a companion paper, I discuss different versions of the Past Hypothesis---the strong, the weak, and the fuzzy---as well as their implications for the uniqueness  of the initial quantum state that we will come to soon. The upshot is that in all cases it will be sufficiently unique  for eliminating statistical mechanical probabilities.   }

In QSM, PH takes the form of \S 2.2 \#9. That is, the microstate (a wave function) starts in a particular low-dimensional subspace in Hilbert space (the PH-subspace). However, it does not pin down a unique microstate. There is still a continuous infinity of possible microstates compatible with the PH-subspace. 

It is plausible to think that, for PH to work as a successful explanation for the Second Law, it has to be on a par with other fundamental laws of nature. That is, we should take PH to be a law of nature and not just a contingent initial condition, for otherwise it might be highly unlikely that our past was in lower entropy and that our inferences to the past are reliable. Already in the context of a weaker version of PH,  \cite{feynman2017character}  suggests that the low-entropy initial condition should be understood as a law of nature. However, PH by itself is not enough. Since there are anti-thermodynamic exceptions even for trajectories starting from the PH-subspace, it is crucial to impose another law about a uniform probability distribution on the subspace.  This is the quantum analog of what Albert (2000) calls the Statistical Postulate (SP). It corresponds to the  measure $\mu$ we specified in  \S 2.2 \#4. We used it to state the typicality statement in \#10. Barry Loewer calls the joint system---the package of laws that includes PH and SP in addition to the dynamical laws of physics---the Mentaculus Vision.\footnote{For developments and defenses of the nomological account of the Past Hypothesis and the Statistical Postulate, see \cite{albert2000time, LoewerCatSLaw, wallace2011logic, wallace2012emergent} and \cite{loewer2016mentaculus}. Albert and Loewer are writing mainly in the context of CSM. The Mentaculus Vision is supposed to provide a ``probability map of the world.'' As such, it requires one to take the probability distribution very seriously. 

To be sure, the view that PH is nomologically fundamental has been debated. See discussions in \cite{price1997time, price2004origins}, and \cite{callender2004measures}. 
However, those challenges are no more threatening to IPH being a law than PH being a law. We will come back to this point after introducing IPH. }

\subsection{Introducing the Initial Projection Hypothesis }

The Past Hypothesis uses a low-entropy macrostate (PH-subspace) to constrain the  microstate  of the system (a state vector in QSM). This is natural from the perspective of Wave Function Realism, according to which the state vector (the wave function) represents the physical degrees of freedom of the system. The  initial state of the system is described by a normalized wave function $\Psi(t_0)$. $\Psi(t_0)$ has to lie in the special low-dimensional Hilbert space $\mathscr{H}_{PH}$ with $dim\mathscr{H}_{PH} \ll dim\mathscr{H}_{eq}$.  Moreover, there are many different choices of initial wave functions in $\mathscr{H}_{PH}$. That is, PH is compatible with many different low-entropy wave functions. Furthermore,  for stating the typicality statements, we also need to specify a measure $\mu$ on the unit sphere of $\mathscr{H}_{PH}$. For the finite-dimensional case, it is just the normalized surface area measure on the unit sphere. 

Density Matrix Realism suggests an alternative way to think about the low-entropy boundary condition.  PH pins down the initial macrostate   $\mathscr{H}_{PH}$, a special subspace of the total Hilbert space. Although $\mathscr{H}_{PH}$ is compatible with many density matrices, there is a natural choice---the normalized projection operator onto $\mathscr{H}_{PH}$. Just as in (\ref{ID}), we can specify it  as: 
\begin{equation}\label{PHID}
\hat{W}_{IPH} (t_0) = \frac{I_{PH}}{dim \mathscr{H}_{PH}},
\end{equation}
where $t_0$ represents a temporal boundary of the universe, $I_{PH}$ is the projection operator onto   $\mathscr{H}_{PH}$, $dim$ counts the dimension of the Hilbert space, and $dim\mathscr{H}_{PH} \ll dim\mathscr{H}_{eq}$. Since the quantum state at $t_0$ has the lowest entropy, we call $t_0$ the initial time.  We shall call (\ref{PHID}) the \emph{Initial Projection Hypothesis} (IPH). In words: the initial density matrix of the universe is the normalized projection onto the PH-subspace.

I propose that we add IPH to any W-quantum theory. The resultant theories will be called $W_{IPH}$-theories.  For example, here are the equations of $W_{IPH}$-BM:
\begin{itemize}
\item[(A)] $\hat{W}_{IPH} (t_0)  = \frac{I_{PH}}{dim \mathscr{H}_{PH}},$

\item[(B)] $P(Q(t_0) \in dq) =  W_{IPH} (q, q, t_0) dq,$

\item[(C)] $i \hbar \frac{\partial \hat{W}}{\partial t} = [\hat{H},  \hat{W}],$

\item[(D)] $\frac{dQ_i}{dt} = \frac{\hbar}{m_i} \text{Im} \frac{\nabla_{q_{i}}  W_{IPH} (q, q', t)}{ W_{IPH} (q, q', t)} (q=q'=Q).$
\end{itemize}
(A) is IPH and (B)---(D) are the defining equations of W-BM. (Given the initial quantum state $\hat{W}_{IPH} (t_0)$, there is a live possibility that for every particle at $t_0$,  its velocity is zero. However, even in this possibility, as long as the initial quantum state ``spreads out'' later, as we assume it would, the particle configuration will typically start moving at a later time.  This is true because of equivariance.\footnote{Thanks to Sheldon Goldstein and Tim Maudlin for discussions here.}) 

Contrast these equations with BM formulated with wave functions and PH (not including SP for now), which will be called  $\Psi_{PH}$-BM:
\begin{itemize}
\item[(A')] $\Psi (t_0)  \in   \mathscr{H}_{PH},$

\item[(B')] $P(Q(t_0) \in dq) =  |\Psi (q, t_0)|^2 dq,$

\item[(C')] $i \hbar \frac{\partial \Psi}{\partial t} = \hat{H}\Psi,$

\item[(D')] $\frac{dQ_i}{dt} = \frac{\hbar}{m_i} \text{Im} \frac{\nabla_{q_{i}}  \Psi(q,t)}{ \Psi(q, t)} (Q).$
\end{itemize}
IPH (A) in $W_{IPH}$-BM plays the same role as PH (A') in $\Psi_{PH}$-BM. Should IPH  be interpreted as a law of nature in  $W_{IPH}$-theories? I think it should be, for the same reason that PH should be interpreted as a law of nature in the corresponding theories. The reason that PH should be interpreted as a law\footnote{See, for example, \cite{feynman2017character, albert2000time, LoewerCatSLaw} and \cite{loewer2016mentaculus}.} is because it is a particularly simple and informative statement that accounts for the widespread thermodynamic asymmetry in time.  PH is simple because it characterizes a simple macrostate $\mathscr{H}_{PH}$, of which the initial wave function is a vector. PH is informative because with PH  the dynamical equations predict time asymmetry and without PH the dynamical equations  cannot. Similarly, IPH is simple because it provides crucial resources for explaining the arrow of time. IPH is informative because it is essential for explaining the time asymmetry in a quantum universe described by a density matrix. (This is in addition to the fact that IPH helps determine the $W_{IPH}$-version of the guidance equation (D).)  To be sure, PH and IPH as laws face the same worries: both are statements about boundary conditions but we usually think of laws as dynamical equations. However, these worries are no more threatening to IPH being a law than PH being a law.

Let us make three remarks about IPH. Firstly, IPH defines a unique initial quantum state. The quantum state $\hat{W}_{IPH} (t_0)$ is informationally equivalent to the constraint that PH imposes on the initial microstates. Assuming that PH selects a unique low-entropy macrostate, $\hat{W}_{IPH} (t_0)$ is singled out by the data in PH.\footnote{The weaker versions of PH are vague about the exact initial low-entropy macrostate. It is vague because, even with a choice of macro-variables, there may be many subspaces that can play the role of a low-entropy initial condition. It would be arbitrary, from the viewpoint of wave-function theories, to pick a specific subspace. In contrast, it would not be arbitrary from the viewpoint of $W_{IPH}$-theories, as the specific subspace defines $W_{IPH}$, which determines the dynamics. }

Secondly, on the universal scale, we do not need to impose an additional probability or typicality measure on the Hilbert space. $\hat{W}_{IPH} (t_0)$ is mathematically equivalent to an integral over projection onto each normalized state vectors (wave functions) compatible with PH \emph{with respect to a normalized surface area measure $\mu$}. But here we are not defining $\hat{W}_{IPH} (t_0)$ in terms of state vectors. Rather, we are thinking of $\hat{W}_{IPH} (t_0)$ as a geometric object in the Hilbert space:  the (normalized) projection operator onto $\mathscr{H}_{PH}$.  That is the \emph{intrinsic} understanding of the density matrix.\footnote{After writing the paper, I discovered that David Wallace has come to a similar idea in a forthcoming paper. There are some subtle differences. He proposes that we can reinterpret  probability distributions in QSM as actual mixed states. Consequently, the problem of statistical mechanical probability is ``radically transformed'' (if not eliminated) in QSM. Wallace's proposal is compatible with different probability distributions and hence different mixed states of the system. It does not require one to choose a particular quantum state such as (\ref{PHID}). Moreover, it is compatible with there being an underlying pure state. In contrast,  I propose a particular, natural initial quantum state of the universe based on the PH subspace---the normalized projection onto the PH subspace (\ref{PHID}), and there is no underlying pure state. As we discuss in \S5.1, this also leads to the elimination of statistical mechanical probability, since the initial state is fixed in the theory. Moreover,  as we discuss below, the natural state inherits the simplicity of the PH subspace, which has implications for the nature of the quantum state. For a more detailed comparison, see \cite{chen2018valia}. }  

Thirdly,  $\hat{W}_{IPH} (t_0)$ is  simple. Related to the first remark, IPH defines $\hat{W}_{IPH} (t_0)$ explicitly as the normalized projection operator onto $\mathscr{H}_{PH}$. There is a natural correspondence between a subspace and its projection operator. If we specify the subspace, we know what its projection operator is, and vice versa. Since the projection operator onto a subspace carries no more information than that subspace itself, the projection operator is no more complex than $\mathscr{H}_{PH}$. This is different from $\Psi_{PH}$, which is confined by PH to be a vector inside $\mathscr{H}_{PH}$. A vector carries more information than the subspace it belongs to, as specifying a subspace is not sufficient to determine  a vector. For example, to determine a vector in an 18-dimensional subspace of a 36-dimensional vector space, we need 18 coordinates in addition to specifying the subspace. The higher the dimension of the subspace, the more information is needed to specify the vector. If PH had fixed  $\Psi_{PH}$ (the QSM microstate), it would have required much more information and  become a much more complex posit. PH as it is  determines $\Psi_{PH}$ only up to an equivalence class (the QSM macrostate).  As we shall see in \S 6, the simplicity of $\hat{W}_{IPH}(t_0)$ will be a crucial ingredient for a new version  of the nomological interpretation of the quantum state.

\subsection{Connections to the Weyl Curvature Hypothesis}

Let us point out some connections between our Initial Projection Hypothesis (IPH) and the Weyl Curvature Hypothesis (WCH) proposed by \cite{penrose1979singularities}. Thinking about the origin of the Second Law of Thermodynamics in the early universe with high homogeneity and isotropy, and the relationship between space-time geometry and entropy, Penrose proposes a low-entropy hypothesis: 
\begin{quote}
I propose, then, that there should be complete lack of chaos in the initial \emph{geometry}. We need, in any case, some kind of low-entropy constraint on the initial state. But thermal equilibrium apparently held (at least very closely so) for the \emph{matter} (including radiation) in the early stages. So the `lowness' of the initial entropy was not a result of some special matter distribution, but, instead, of some very special initial spacetime geometry. The indications of [previous sections], in particular, are that this restriction on the early geometry should be something like: \emph{the Weyl curvature $C_{abcd}$ vanishes at any initial singularity}. (\cite{penrose1979singularities}, p.630, emphasis original)
\end{quote}
The Weyl curvature tensor $C_{abcd}$ is the traceless part of the Riemann curvature tensor $R_{abcd}$. It is not fixed completely by the stress-energy tensor and thus has independent degrees of freedom in Einstein's general theory of relativity. Since the entropy of the matter distribution is quite high, the origin of thermodynamic asymmetry should be due to the low entropy in geometry, which corresponds very roughly to the vanishing of the Weyl curvature tensor. 

WCH is an elegant and simple way of encoding the initial low-entropy boundary condition in the classical spacetime geometry. If WCH could be extended to a quantum theory of gravity, presumably it  would pick out a simple subspace (or subspaces) of the total Hilbert space that corresponds to $C_{abcd} \rightarrow 0$. Applying IPH to such a theory, the initial density matrix will be the normalized projection onto that subspace (subspaces).\footnote{There is another connection between the current project and Penrose's work. The W-Everettian theory that we considered in \S 3.2 combined with the Initial Projection Hypothesis is a theory that satisfies \emph{strong determinism} (\cite{roger1989emperor}). This is because the entire history of the W$_{IPH}$-Everettian universe described by  $W_{IPH}(t)$, including its initial condition, is fixed by the laws. }

\section{Theoretical Payoffs}

 W$_{IPH}$-quantum theories, the result of applying IPH to W-theories, have two  theoretical payoffs, which we  explore in this section. These are by no means decisive arguments in favor of the density-matrix framework, but they display some interesting differences with the wave-function framework. 

\subsection{Harmony between Statistical Mechanics and Quantum Mechanics}

In W$_{IPH}$-quantum theories, statistical mechanics is made more harmonious with quantum mechanics.  As we pointed out earlier, standard QM and QSM contain the wave function in addition to the density matrix, and they require the addition of both the Past Hypothesis (PH) and the Statistical Postulate (SP) to the dynamical laws. In particular, we have two kinds of probabilities: the quantum-mechanical ones (Born rule probabilities) and the statistical mechanical ones (SP). The situation is quite different in our framework.  This is true for all the W$_{IPH}$-theories. We will use W$_{IPH}$-BM ((A)---(D)) as an example.

 W$_{IPH}$-BM completely specifies the initial quantum state, unlike $\Psi_{PH}$-BM. For $\Psi_{PH}$-BM, because of time-reversal invariance, some initial wave functions compatible with PH will evolve to lower entropy. These are called  \emph{anti-entropic} exceptions. However, the uniform probability distribution (SP) assigns low probability to these exceptions. Hence, we expect that with overwhelming probability the actual wave function is entropic. For W$_{IPH}$-BM, in contrast, there is no need for something like SP, as there is only one initial density matrix compatible with IPH---$W_{IPH} (t_0)$. It is guaranteed to evolve to future states that have entropic behaviors. Therefore, on the universal scale,   W$_{IPH}$-BM   eliminates the need for SP and thus the need for a  probability/typicality measure that is in addition to the quantum-mechanical measure (B). This is a nice feature of W$_{IPH}$-theories, as it is desirable to unify the two sources of randomness: quantum-mechanical and statistical-mechanical.  Of course, wave functions and statistical-mechanical probabilities are still useful to analyze subsystems such as gas in a box, but they no longer play fundamental roles in W$_{IPH}$-theories. Another strategy to eliminate SP has been explored in the context of GRW jumps by \cite{albert2000time}. \cite{wallace2011logic, wallace2012emergent} has proposed a replacement of SP with a non-probabilistic constraint on the microstate, giving rise to the \emph{Simple Dynamical Conjecture}. These are quite different proposals, all of which deserve further developments.

\subsection{Descriptions of the Universe and the Subsystems}

W$_{IPH}$-quantum theories also bring more unity to the kinematics and the dynamics of the universe and the subsystems. 

Let us start with a quantum-mechanical universe $U$.  Suppose it contains many subsystems. Some of them will be interacting heavily with the environment, while others will be effectively isolated from the environment. 
For a universe that contains some quasi-isolated subsystems (interactions with the environment effectively vanish), the following is a desirable property: 
\begin{description}
\item \textsc{Dynamic Unity} The dynamical laws of the universe are the same as the effective laws of most quasi-isolated subsystems. 
\end{description} 
Dynamic Unity is a property that can come in degrees, rather than an ``on-or-off'' property. Theory A  has more dynamic unity than Theory B, if the fundamental equations in A are valid in more subsystems than those in B. This property is desirable, but not indispensable. It is desirable because  law systems that apply both at the universal level and at the subsystem level are unifying and explanatory. 


W-BM has more dynamic unity than BM formulated with a universal wave function. 
For quantum systems without spin, we can always follow \cite{durr1992quantum} to define \emph{conditional wave functions} in BM. For example, if the universe is partitioned into a system $S_1$ and its environment $S_2$, then for $S_1$, we can define its conditional wave function:
\begin{equation}
\psi_{cond}(q_1) = C \Psi(q_1, Q_2),
\end{equation}
where C is a normalization factor and $Q_2$ is the actual configuration of $S_2$. $\psi_{cond}(q_1)$ always gives the velocity field for the particles in $S_1$ according to the guidance equation. However, for quantum systems with spin, this is not always true. Since BM is described by ($\Psi(t), Q(t)$), it does not contain actual values of spin. Since there are no actual spins to plug into the spin indices of the wave function, we cannot always define conditional wave functions in an analogous way. Nevertheless, in those circumstances, we can follow \cite{durr2005role} to define a \emph{conditional density matrix} for $S_1$, by plugging in the actual configuration of $S_2$ and tracing over the spin components in the wave function associated with $S_2$.\footnote{ The conditional density matrix for $S_1$ is defined as:
\begin{equation}\label{WCond}
{W_{cond}}_{s_1'}^{s_1}(q_1, q_1') = \frac{1}{N} \sum_{s_2} \Psi^{s_1 s_2} (q_1, Q_2) \Psi_{s_1 s_2}^* (q_1', Q_2),
\end{equation}
with the normalizing factor:
\begin{equation}\label{Norm}
N = \int_{\mathcal{Q}_1} dq_1 \sum_{s_1 s_2} \Psi^{s_1 s_2} (q_1, Q_2) \Psi_{s_1 s_2}^* (q_1', Q_2).
\end{equation}
} 
The conditional density matrix will guide the particles in $S_1$ by the W-guidance equation (the spin version with the partial trace operator). 

In W-BM, the W-guidance equation is always valid for the universe and the subsystems. In BM, sometimes subsystems do not have conditional wave functions, and thus the wave-function version of the guidance equation is not always valid. In this sense, the W-BM equations are valid in more circumstances than the BM equations. However, this point does not rely on IPH. 

What about Everettian and GRW theories? Since  GRW and Everettian theories do not have fundamental particles, we cannot obtain conditional wave functions for subsystems as in BM. However, even in the $\Psi$-versions of GRW and Everett, many subsystems will not have pure-state descriptions by wave functions due to the prevalence of entanglement. Most subsystems can be described only by a mixed-state density matrix, even when the universe as a whole is described by a wave function. In contrast, in W$_{IPH}$-Everett theories and W$_{IPH}$-GRW theories, there is more uniformity across the subsystem level and the universal level: the universe as a whole as well as most subsystems are described by the same kind of object---a (mixed-state) density matrix. Since state descriptions concern the kinematics of a theory, we say that W-Everett and W-GRW theories have more \emph{kinematic unity} than their $\Psi$-counterparts:
\begin{description}
\item \textsc{Kinematic Unity} The state description of the universe is of the same kind as the state descriptions of most quasi-isolated subsystems. 
\end{description}



So far, my main goal has been to show that Density Matrix Realism + IPH  is a viable position. They have theoretical payoffs that are interestingly different from those in the original package (Wave Function Realism + PH).  In the next section, we look at their relevance to the nature of the quantum state.

\section{The Nomological Thesis}

Combining Density Matrix Realism with IPH gives us W$_{IPH}$-quantum theories that have interesting theoretical payoffs. We have also argued that the initial quantum state in such theories would be simple and unique. In this section, we show that the latter fact lends support to the nomological interpretation of the quantum state:
\begin{description}
\item[The Nomological Thesis:] The initial quantum state of the world is nomological.
\end{description}
However, ``nomological'' has several senses and has been used in several ways in the literature. We will start with some clarifications. 

\subsection{The Classical Case}

We can clarify the sense of the``nomological'' by taking another look at classical mechanics. In classical $N$-particle Hamiltonian mechanics, it is widely accepted that the Hamiltonian function is nomological, and that the ontology consists in particles with positions and momenta. Their state is given by $X=(\boldsymbol{q_1}(t), ..., \boldsymbol{q_N}(t); \boldsymbol{p_1}(t),...,\boldsymbol{p_n}(t))$, and the Hamiltonian is $H=H(X)$. Particles move according to the Hamiltonian equations:
\begin{equation}\label{HE}
\frac{d \boldsymbol{q_i}(t)}{d t} = \frac{\partial H}{\partial \boldsymbol{p_i}} \text{  ,  } \frac{d \boldsymbol{p_i}(t)}{d t} = - \frac{\partial H}{\partial \boldsymbol{q_i}}.
\end{equation}
 Their motion corresponds to a trajectory in  phase space. The velocity field on  phase space is obtained by taking suitable derivatives of the Hamiltonian function $H$. The equations have the form: 
\begin{equation}\label{N1}
\frac{dX}{dt} = F (X) = F^H (X) 
\end{equation}
Here, $F^H(X)$ is $H(q,p)$ with suitable derivative operators. The  Hamiltonian equations have a simple form, because $H$ is simple. $H$ can be written explicitly as follows:
\begin{equation}\label{HF}
H = \sum^{N}_{i} \frac{p_i^2}{2m_i} + V,
\end{equation}
where $V$ takes on this form when we consider electric and gravitational potentials:
\begin{equation}\label{PE}
V = \frac{1}{4\pi \epsilon_0} \sum_{1\leq j\leq k\leq N} \frac{e_j e_k}{|q_j - q_k|} + \sum_{1\leq j\leq k\leq N} \frac{G m_j m_k}{|q_j - q_k|} ,
\end{equation}
That is, the RHS of the Hamiltonian equations,  after making the Hamiltonian function explicit,  are still simple. $H$ is just a convenient shorthand for (\ref{HF}) and (\ref{PE}). Moreover, $H$  is also fixed by the theory. A classical universe is governed by the dynamical laws plus the fundamental interactions. If $H$ were different in (\ref{N1}), then we would have a different physical theory (though it would still belong to the class of theories called classical mechanics). For example, we can add another term in (\ref{PE}) to encode another fundamental interaction, which will result in a different theory. 

Consequently, it is standard to interpret $H$ as a function in (\ref{HE}) that does not represent things or properties of the ontology. Expressed in terms of $H$, the equations of motion take a particularly simple form. The sense that $H$ is nomological is that (i) it generates motion, (ii) it is simple,  (iii) it is fixed by the theory (nomologically necessary), and (iv) it does not represent things in the ontology.  In contrast, the position and momentum variables in (\ref{HE}) are ``ontological'' in that they represent things and properties of the ontology,  take on complicated values, change according to $H$, and are not completely fixed by the theory (contingent). 

\subsection{The Quantum Case}

It is according to the above sense that \cite{goldstein1996bohmian, goldstein2001quantum}, and \cite{goldstein2013reality}  propose that the universal wave function in BM is nomological (and governs things in the ontology). With the guidance equation, $\Psi$ generates the motion of particles. It is of the same form as above:
\begin{equation}\label{N2}
\frac{dX}{dt} = F (X) = F^{\Psi} (X). 
\end{equation}
Why is it simple? Generic wave functions are not simple. However, they observe that, in some formulations of quantum gravity, the universal wave function satisfies the Wheeler-DeWitt equation and is therefore stationary. To be stationary, the wave function does not have time-dependence and probably has many symmetries, in which case it \emph{could} be quite simple. The Bohmian theory then will explicitly stipulate what the universal wave function is. Therefore, in these theories,  provided that $\Psi$ is sufficiently simple, we can afford the same interpretation of  $\Psi$ as we can for $H$ in classical mechanics: both are nomological in the above sense. 

$W_{IPH}$-BM  also supports the nomological interpretation of the quantum state but via a different route. With the W-guidance equation, $W_{IPH}$ generates the motion of particles. It is of the same form as above:
\begin{equation}\label{N3}
\frac{dX}{dt} = F (X) = F^{W_{IPH}} (X). 
\end{equation}
Why is it simple? Here we do not need to appeal to specific versions of quantum gravity, which are still to be worked out and may not guarantee the simplicity of $\Psi$. Instead, we can just appeal to IPH. We have argued in \S4.2 that IPH is simple and that $W_{IPH}(t_0)$ is simple. Since the quantum state evolves unitarily by the von Neumann equation, we can obtain the quantum state at any later time as:
\begin{equation}\label{Wt}
\hat{W}_{IPH} (t) = e^{-i\hat{H}t/\hbar} \hat{W}_{IPH} (t_0) e^{i\hat{H}t/\hbar}
\end{equation}
Since $W_{IPH}(t)$ is a simple function of the time-evolution operator and the initial density matrix, and since both are simple, $W_{IPH}(t)$ is also simple. So we can think of $W_{IPH}(t)$ just as a convenient shorthand for (\ref{Wt}). (This is not true for $\ket{\Psi(t)} = \hat{H} \ket{\Psi(t_0)}$, as generic $\ket{\Psi(t_0)}$ is not simple at all.)

The ``shorthand'' way of thinking about $W_{IPH}(t)$ implies that  the equation of particle motion has a time-dependent form $F^{W_{IPH}} (X, t)$. Does time-dependence undercut the nomological interpretation? It does not in this case, as the $F^{W_{IPH}} (X, t)$ is still simple even with time-dependence. It is true that time-independence is often a hallmark of a nomological object, but it is not always the case. In this case, we have simplicity without time-independence. Moreover, unlike the proposal of \cite{goldstein1996bohmian, goldstein2001quantum}, and \cite{goldstein2013reality}, we do not need time-independence to argue for the simplicity of the quantum state. 

Since $W_{IPH}(t_0)$ is fixed by IPH, $F^{W_{IPH}}$ is also fixed by the theory. Let us expand (\ref{N3}) to make it more explicit:
\begin{equation}\label{N4}
\frac{dQ_i}{dt} = 
\frac{\hbar}{m_i} \text{Im} \frac{\nabla_{q_{i}}   W_{IPH} (q, q', t)}{W_{IPH} (q, q', t)} (Q) = 
\frac{\hbar}{m_i} \text{Im} \frac{\nabla_{q_{i}}   \bra{q} e^{-i \hat{H} t/\hbar} \hat{W}_{IPH} ( t_0) e^{i \hat{H} t/\hbar} \ket{q'} }{ \bra{q} e^{-i \hat{H} t/\hbar} \hat{W}_{IPH} (t_0) e^{i \hat{H} t/\hbar} \ket{q'}} (q=q'=Q)
\end{equation}
The initial quantum state (multiplied by the time-evolution operators) generates motion, has a simple form, and is fixed by the boundary condition (IPH) in $W_{IPH}$-BM. Therefore, it is nomological. This is of course a modal thesis. The initial quantum state, which is completely specified by IPH, could not have been different.

Let us consider other $W_{IPH}$-theories with local beables. In $W_{IPH}$-Sm, the initial quantum state has the same simple form and is fixed by IPH. It does not generate a velocity field, since there are no fundamental particles in the theory. Instead, it determines the configuration of the mass-density field on physical space. This is arguably different from the sense of nomological that $H$ in classical mechanics displays. Nevertheless, the mass-density field and the Bohmian particles play a similar role---they are ``local beables'' that make up tables and chairs, and they are governed by the quantum state. In $W_{IPH}$-GRWm and $W_{IPH}$-GRWf, the initial quantum state has the same simple form and is fixed by IPH. It does not generate a velocity field, and it evolves stochastically. This will determine a probability distribution over configurations of local beables---mass densities or flashes---on physical space. The initial quantum state in these theories can be given an \emph{extended nomological interpretation}, in the sense that condition (i) is extended such that it covers other kinds of ontologies and dynamics: (i') the quantum state determines (deterministically or stochastically) the configuration of local beables. 

The $W_{IPH}$-theories with local beables support the nomological interpretation of the initial quantum state. It can be interpreted in non-Humean ways and Humean ways. On the non-Humean proposal, we can think of the initial quantum state as an additional nomological entity that \emph{explains} the distribution of particles, fields, or flashes. On the Humean proposal, in contrast, we can think of the initial quantum state as something that \emph{summarizes} a separable mosaic. This leads to reconciliation between Humean supervenience and quantum entanglement. 

\subsection{Humean Supervenience}

Recall that according to Humean supervenience (HS), the ''vast mosaic of local matters of particular fact'' is a \emph{supervenience base} for everything else in the world, the \emph{metaphysical ground floor} on which everything else depends. On this view, laws of physics are nothing over and above the ``mosaic.'' They are just the axioms in the simplest and most informative summaries of the local matters of particular fact. A consequence of HS is that the complete physical state of the universe is determined by the properties and spatiotemporal arrangement of the local matters (suitably extended to account for vector-valued magnititudes) of particular facts. It follows that there should not be any state of the universe that fails to be determined by the properties of individual space-time points.\footnote{This is one reading of David Lewis. Tim Maudlin (\citeyear{MaudlinMWP}) calls this thesis ``Separability.''} Quantum entanglement, if it were in the fundamental ontology, would present an obstacle to HS, because entanglement is not determined by the properties of space-time points. 
The consideration above suggests a strong  \emph{prima facie} conflict between HS and quantum physics. On the basis of  quantum non-separability, Tim Maudlin has proposed an influential argument against HS.\footnote{See \cite{MaudlinMWP}, Chapter 2.} 

$W_{IPH}$-theories with local beables offer a way out of the conflict between quantum entanglement and Humean supervenience. A Humean can interpret the laws  (including the IPH) as the axioms in the best system that summarize a separable mosaic. Take W$_{IPH}$-BM as an example:

\begin{description}
\item[The W$_{IPH}$-BM mosaic:] particle trajectories $Q(t)$ on physical space-time.

\item[The W$_{IPH}$-BM best system:]  four equations---the simplest and strongest axioms summarizing the mosaic: 

\begin{itemize}
\item[(A)] $\hat{W}_{IPH} (t_0)  = \frac{I_{PH}}{dim \mathscr{H}_{PH}}$

\item[(B)] $P(Q(t_0) \in dq) =  W_{IPH} (q, q, t_0) dq,$

\item[(C)] $i \hbar \frac{\partial \hat{W}}{\partial t} = [\hat{H},  \hat{W}],$

\item[(D)] $\frac{dQ_i}{dt} = \frac{\hbar}{m_i} \text{Im} \frac{\nabla_{q_{i}}   W_{IPH} (q, q', t)}{W_{IPH} (q, q', t)} (q=q'=Q).$
\end{itemize}
\end{description}
Notice that (A)---(D) are  simple and informative statements about $Q(t)$. They are expressed in terms of  $\hat{W}_{IPH} (t)$, which via law (C) can be expressed in terms of $\hat{W}_{IPH} (t_0)$. We have argued previously that the initial quantum state can be given a nomological interpretation. The Humean maneuver is that the law statements are to be understood as axioms of the best summaries  of the mosaic. The mosaic described above is completely separable, while the best system,  completely  specifying the quantum state and the dynamical laws, contains all the information about quantum entanglement and superpositions. The entanglement facts are no longer fundamental. As on the original version of Humean supervenience, the best system consisting of (A)---(D) supervenes on the mosaic. Hence, this proposal reconciles Humean supervenience with quantum entanglement. As it turns out, the above version of Quantum Humeanism also achieves more theoretical harmony, dynamical unity, and kinematic unity (\S 5), which are desirable from the Humean best-system viewpoint.  We can perform similar ``Humeanization'' maneuvers on the density matrix in other quantum theories with local beables---W-GRWm, W-GRWf, and W-Sm (although such procedures might not be as compelling).

This version of Quantum Humeanism based on $W_{IPH}$-theories is different from the other approaches in the literature: \cite{AlbertEQM, LoewerHS, miller2014quantum, esfeld2014quantum, bhogal2015humean, callender2015one} and \cite{esfeld2017minimalist}. In contrast to the high-dimensional proposal of \cite{AlbertEQM} and \cite{LoewerHS}, our version preserves the fundamentality of physical space. 

The difference between our version and those of \cite{miller2014quantum, esfeld2014quantum, bhogal2015humean, callender2015one}, and \cite{esfeld2017minimalist} is more subtle.  They are concerned primarily with $\Psi$-BM. We can summarize their views as follows (although they do not agree on all the details). There are several parts to their proposals. First, the wave function is merely part of the best system. It is more like parameters in the laws such as mass and charge. Second, just like the rest of the best system, the wave function supervenes on the mosaic of particle trajectories. Third, the wave function does not have to be very simple. The Humean theorizer, on this view, just needs to find the simplest and strongest summary of the particle histories, but the resultant system can  be  complex \emph{simpliciter}. One interpretation of this view is that the best system for $\Psi_{PH}$-BM is just (A')---(D') in \S4.2 (although they do not explicitly consider (A')), such that neither the mosaic nor the best system specifies the exact values of the universal wave function. In contrast, our best system completely specifies the universal quantum state. The key difference between our approaches is that their interpretation of the wave function places much weaker constraints than our nomological interpretation does. It is much easier for something to count as being part of the best system on their approach than on ours. While they do not require the quantum state to be simple, we do. For them, the Bohmian guidance equation is likely very complex after plugging in the actual wave function $\Psi_{PH}$ on the RHS, but $\Psi_{PH}$ can still be part of their best system.\footnote{See \cite{DewarHS} \S5 for some worries about the weaker criterion on the best system.} For us, it is crucial that the equation remains simple after plugging in $W_{IPH} (t_0)$ for it to be in the best system. Consequently, $W_{IPH}(t_0)$ is nomological in the sense spelled out in \S6.1, and we can give it a Humean interpretation similar to that of the Hamiltonian function in CM. Generic $\Psi_{PH}$, on the other hand, cannot be nomological in our sense. But that is ok for them, as their best-system interpretation does not require the strong nomological condition that we use. Here we do not attempt to provide a detailed comparison; we   do that in  \cite{chen2018HU}.

\section{Conclusion}

I have introduced a new package of views: Density Matrix Realism, the Initial Projection Hypothesis, and the Nomological Thesis. In the first two steps, we introduced a new class of quantum theories---W$_{IPH}$-theories. In the final step, we argue that it is a theory in which the initial quantum state \emph{can} be given a nomological interpretation. Each is interesting in its own right, and they do not need to be taken together.  However,  they fit together quite well. They provide alternatives to standard versions of realism about quantum mechanics, a new way to get rid of statistical-mechanical probabilities,  and a new solution to the conflict between quantum entanglement and Humean Supervenience. To be sure, there are many other features of W$_{IPH}$-theories in general and the nomological interpretation in particular that are worth exploring further. 

The most interesting feature of the new package, I think, is that it brings together the foundations of quantum mechanics and quantum statistical mechanics. In W$_{IPH}$-theories,  the arrow of time becomes intimately connected to the quantum-mechanical phenomena in nature. It is satisfying to see that nature is so unified.


\section*{Acknowledgement}

I would like to thank  the editors and referees of \emph{The British Journal for the Philosophy of Science} for helpful feedback. I am also grateful for stimulating discussions with   David Albert, Sheldon Goldstein, and Barry Loewer. I have received helpful feedback from 
Abhay Ashtekar, Karen Bennett, Harjit Bhogal, Max Bialek, Miren Boehm, Robert Brandenberger, Tad Brennan, Craig Callender, Sean Carroll, Eugene Chua, Juliusz Doboszewski, Detlef D\"urr, Denise Dykstra, Michael Esfeld, Veronica Gomez, Hans Halvorson, Harold Hodes, Mario Hubert, Michael Kiessling, Dustin Lazarovici, Stephen Leeds, Matthias Lienert, Niels Linnemann, Chuang Liu, Vera Matarese, Tim Maudlin, Kerry McKenzie, Elizabeth Miller, Sebastian Murgueitio, Wayne Myrvold, Jill North, Zee Perry, Davide Romano,  Ezra Rubenstein, Charles Sebens, Jonathan Schaffer, Ted Sider, Joshua Spencer, Noel Swanson, Karim Th\'ebault, Anncy Thresher, Roderich Tumulka, David Wallace, Isaac Wilhelm, Nino Zangh\`i, audiences at Cornell University,  University of Western Ontario,  University of Wisconsin-Milwaukee,  University of Wisconsin-Madison,  University of California, San Diego, and the 2018 Rotman Summer Institute in Philosophy of Cosmology.



\bibliography{test}


\end{document}